\documentclass[journal,twoside,web]{ieeecolor}
\usepackage{generic}
\usepackage{cite}
\usepackage{amsmath,amssymb,amsfonts}
\usepackage{graphicx}
\usepackage{dblfloatfix}
\usepackage{textcomp}
\usepackage{booktabs}
\usepackage{tabularx}
\usepackage{multirow}
\usepackage{array}
\usepackage{url}
\usepackage{hyperref}
\hypersetup{hidelinks=true}

\def\BibTeX{{\rm B\kern-.05em{\sc i\kern-.025em b}\kern-.08em
    T\kern-.1667em\lower.7ex\hbox{E}\kern-.125emX}}
\begin{document}

\title{ImProNCDE: Impulse-Corrected Neural Controlled Differential Equations with Prototype Learning for Longitudinal Prognosis Prediction}

\author{Hao Wang, Yupeng Xu, Jinghao Lin, Shuchang Ye, Yige Peng, Jinman Kim, Kun Liu, and Lei Bi
\thanks{Hao Wang, Yupeng Xu, and Jinghao Lin contributed equally to this work. Corresponding authors: Jinman Kim, Kun Liu, and Lei Bi.}
\thanks{Hao Wang, Yige Peng, and Lei Bi are with Shanghai Jiao Tong University, Shanghai, China (e-mail: lei.bi@sjtu.edu.cn).}
\thanks{Yupeng Xu and Kun Liu are with Shanghai General Hospital, Shanghai Jiao Tong University School of Medicine, Shanghai, China (e-mail: drliukun@sjtu.edu.cn).}
\thanks{Jinghao Lin is with Northeastern University, Shenyang, Liaoning, China.}
\thanks{Shuchang Ye and Jinman Kim are with The University of Sydney, Sydney, NSW, Australia (e-mail: jinman.kim@sydney.edu.au).}}

\maketitle
\begin{abstract}
Longitudinal ophthalmic imaging analysis is an essential step for prognosis prediction in ophthalmic diseases. However, AI-assisted prognosis models are challenged by follow-up sequences, which tend to be sparse, irregularly sampled, and incomplete. Although advanced prognosis modeling methods, especially for the methods based on neural controlled differential equations (NCDEs), provide a principled continuous-time framework for sparse and irregular longitudinal data. Unfortunately, two major concerns remain unsolved in clinical follow-up modeling. First, the smooth latent dynamics of standard NCDEs is poorly matched to abrupt pathological changes induced by therapeutic intervention, lesion recurrence, or long follow-up gaps. Second, numerical integration over long horizons can accumulate errors, which will produce unstable latent trajectories and weakened class discrimination. To address these challenges, we propose ImProNCDE, an impulse-corrected NCDE framework with prototype learning for longitudinal ophthalmic prognosis prediction. To capture abrupt pathological changes beyond smooth latent dynamics, ImProNCDE introduces Residual Impulse Calibration (RIC), which injects residual-based impulse corrections at visit times and then recalibrates the latent state when observations deviate from continuous predictions. To further mitigate error accumulation over long horizons, we introduce a Prototype-guided Trajectory Stabilizer (PTS), which aims to attract latent trajectories toward learnable prognosis prototypes to reduce class overlap and which ultimately improves long-horizon stability. Experiments on multiple private and public longitudinal ophthalmic datasets (totalling over 1206 samples) show that ImProNCDE outperforms existing SOTA methods focusing on sequence modeling.
\end{abstract}


\begin{figure}[t!]
    \centering
    \includegraphics[width=\columnwidth]{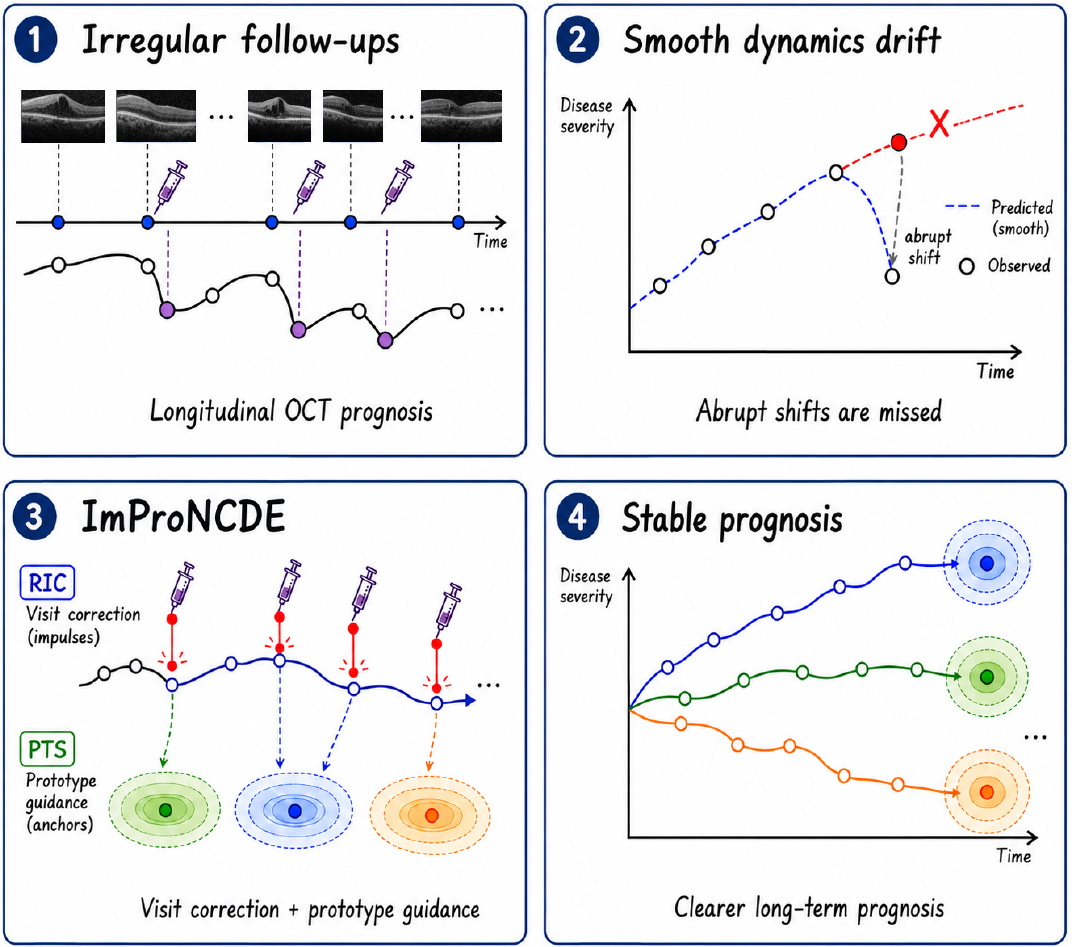}
    \caption{Overview of ImProNCDE for longitudinal ophthalmic prognosis prediction. Existing longitudinal models fail to capture abrupt pathological transitions during irregular OCT follow-ups, causing latent trajectories to drift from the observed pathological state. ImProNCDE mitigates this problem through residual impulse calibration and prototype-guided trajectory stabilization, as such enable more stable long-term prognosis prediction.}
    \label{fig:overview}
\end{figure}

\section{Introduction}
\IEEEPARstart{L}{ongitudinal} prognostic prediction provides clinically useful guidance for personalized treatment planning in AI-assisted ophthalmology, supporting precision medicine by guiding follow-up decisions and reducing unnecessary interventions.\cite{miguel2022oct,suda2018evaluation,liu2022prospective,chen2025radiomics}. For many ophthalmic diseases, treatment management heavily relies on repeated follow-up, where clinicians use serial ophthalmic images to assess anatomical response, monitor disease activity, and adjust treatment schedules \cite{brown2022kestrel,singh2023efficacy,murata2025predictive}. Accurate prognosis from these longitudinal imaging sequences can help to distinguish patients who are likely to respond well from those at risk of persistent or recurrent disease, which thereby being able to support more precise follow-up planning and reducing unnecessary treatment burden. However, follow-up imaging is often affected by disease status and patient adherence, resulting in clinical imaging sequences are sparse and irregularly sampled \cite{okudan2025optical}. Therefore, an accurate prognostic model could capture deviations in pathological trajectories during clinical follow-up and then accurately predict prognostic outcomes is highly desirable.

Existing methods try to estimate longitudinal prognosis from two perspectives. The first perspective treats prognosis as temporal sequence classification, where RNNs, LSTMs, and transformers aggregate features derived at individual time-point visit to predict the final outcome \cite{elman1990finding,hochreiter1997long,vaswani2017attention}. Recent longitudinal medical image models further improve this paradigm with time-aware representations, temporal augmentation, or continuous-attention designs \cite{chen2023contiformer,sun2024lomia}. However, these methods typically require additional time-aware modeling to embed temporal information, which limits their ability to naturally capture continuous pathological evolution between follow-ups. As a result, sparse and irregular follow-up intervals are difficult to represent. The second perspective predicts future images or future latent states and then infers disease status from the predicted follow-up information. Diffusion-based methods are the representative methods for this direction as they can model conditional denoising, medical classification, sequence-aware generation, multimodal prediction, and treatment-aware longitudinal synthesis \cite{han2022card,yang2023diffmic,yoon2023sadm,wu2024mmfusion,liu2025treatment}. Nevertheless, their prognostic reliability is sensitive to generation or denoising quality, and prediction may become inaccurate when unobserved events between visits change the pathological trajectory. In addition, diffusion-style models often depend on large-scale pretraining, which is difficult to be adapted for longitudinal ophthalmic imaging cohorts in a specific disease type or modalities.

To better describe disease changes across uneven follow-up intervals without explicitly generating future images, neural differential equation models represent progression in continuous time. Earlier clinical temporal models attempted to handle irregularity by adding elapsed time or missing patterns to sequence encoders \cite{baytas2017patient,che2018recurrent}, while ODE-RNN and Latent ODE moved the hidden state itself into a continuous-time dynamical system \cite{rubanova2019latent}. Neural ordinary differential equations (NODEs) further formalize this idea by parameterizing the derivative of a latent state with a neural vector field and obtaining the next state through numerical integration \cite{chen2018neural,wu2022nodeo,xu2021multi}. However, a standard NODE trajectory is mainly determined by the latent state and vector field, so it does not directly use the full observation stream to control the evolution after initialization. Neural controlled differential equations (NCDEs) address this issue by constructing an interpolated observation path as the control signal, as such allowing latent trajectories to be continuously driven by irregularly sampled measurements \cite{kidger2020neural}. These developments show that observation-driven continuous dynamics are well suited to sparse clinical follow-up data because the model can evolve between visits while still being guided by the measured patient trajectory.

Despite their advantages, existing NCDE-based methods still leave two key problems unresolved for longitudinal ophthalmic prognosis. First, disease evolution usually follows a hybrid trajectory, where gradual pathological progression is interrupted by abrupt anatomical changes caused by therapeutic intervention, lesion recurrence, or shifts in disease activity. As a consequence, long intervals between visits can therefore create large discrepancies between the continuous trajectory predicted from previous observations and the newly observed characteristics. Standard NCDEs continuously evolve the latent state, but they do not explicitly recalibrate the state when a new visit reveals such trajectory deviations. Second, when abrupt pathological shifts are not explicitly captured during continuous-time integration, the solver continues to propagate the latent state along an outdated smooth trajectory. The resulting mismatch can accumulate over long prognostic horizons, which leads the endpoint representation to deviate from the original pathological status and toward an incorrect prognosis.

To bridge these gaps, we propose ImProNCDE, an impulse-corrected NCDE framework with prototype learning for modeling irregular longitudinal ophthalmic imaging trajectories and to abrupt pathological shifts during clinical follow-up. ImProNCDE uses a pretrained retinal foundation encoder to extract visit-level imaging embeddings and then lifts them into a continuous control path for NCDE-based trajectory modeling. To handle follow-up trajectory deviations revealed by newly observed visits, we introduce Residual Impulse Calibration (RIC), which compares the NCDE-predicted state with the observed imaging representation and injects a residual-based correction into the latent trajectory. To reduce long-horizon drift, we further introduce a Prototype-guided Trajectory Stabilizer (PTS), which uses learnable prognosis prototypes as class-aware anchors for the evolving latent state. Experiments on longitudinal ophthalmic datasets show that ImProNCDE achieves strong performance when compared with traditional sequence models, medical time-series backbones, and adapted diffusion-based prognosis backbones. Our core contributions are summarized as follows:

\noindent \textbf{1)} We propose \textbf{ImProNCDE}, an impulse-corrected NCDE framework for irregular longitudinal ophthalmic prognosis prediction. We purposely designed ImProNCDE to target sparse and irregular intervention-associated follow-up trajectories, where ImProNCDE automatically captures discontinuous pathological progression through continuous-time dynamics and then calibrates deviated disease evolution trajectories in a prototype space, as such enables more accurate prognosis prediction for patients with sparse and irregular follow-ups.

\noindent \textbf{2)} We develop \textbf{Residual Impulse Calibration (RIC)} to recalibrate latent trajectories at observed follow-up visits. When compared with existing methods, where newly acquired imaging features deviate from the latent states tend to lead to inaccurate prognosis prediction, RIC measures this residual mismatch and then applies an adaptive impulse correction to realign the latent trajectory with the observed data.

\noindent \textbf{3)} We develop a \textbf{Prototype-guided Trajectory Stabilizer (PTS)} to address long-horizon latent drift under sparse follow-up observations. PTS uses learnable prognosis prototypes and Mahalanobis geometry to constrain corrected trajectories toward class-aware latent regions, thereby improving endpoint stability and prognosis separability.

\section{Related Work}

\subsection{Longitudinal Prognosis Prediction in Ophthalmology}
Ophthalmic prognosis research has gradually moved from disease recognition toward treatment-response and long-term outcome prediction. Early ophthalmic imaging studies showed that structural biomarkers and quantitative imaging measurements can support anti-VEGF demand estimation, visual outcome prediction, retreatment interval assessment, and long-term retinal disease monitoring \cite{miguel2022oct,suda2018evaluation,gallardo2021machine,bogunovic2022predicting}. This direction was further extended by models that combine raw ophthalmic images with demographic or clinical variables to predict treatment outcome in neovascular age-related macular degeneration and related retinal diseases \cite{yeh2022prediction,murata2025predictive}. These studies established an important clinical premise: longitudinal ophthalmic imaging contains prognostic signals that can guide follow-up planning, treatment burden estimation, and individualized management.

Recent deep learning studies have made this premise more explicit by using serial images rather than only baseline measurements. For example, recurrence and treatment-response prediction models use multiple imaging visits during the loading or early treatment phase to estimate future disease activity \cite{jung2024prediction,han2024anti}, while broader longitudinal imaging frameworks and radiomics-style analyses explore long-term progression risks from ophthalmic image-derived features \cite{holste2024harnessing,chen2025radiomics}. Multimodal prognosis studies further incorporate clinical variables for individualized anti-VEGF decision support in diabetic macular edema and other retinal settings \cite{mondal2025application}. Nevertheless, existing ophthalmic prognosis methods remain fragmented across diseases, endpoints, and visit designs, and many of them still rely on engineered biomarkers, fixed observation windows, or sequence aggregation without explicitly modeling irregular continuous disease trajectories. When applied to irregular longitudinal ophthalmic prognosis, these limitations can obscure abrupt trajectory deviations caused by treatment response or recurrent disease activity, leading to less reliable endpoint prediction.

\subsection{Temporal Models for Irregular Clinical Sequences}
Temporal modeling methods provide the algorithmic foundation for learning disease progression from sequential clinical observations. Classical recurrent models such as RNNs and LSTMs represent visit order through hidden-state propagation and have been widely used for sequence classification \cite{elman1990finding,hochreiter1997long}. Time-aware variants further incorporate elapsed time, missingness patterns, or decay mechanisms into the sequence encoder, enabling clinical records with irregular measurements to be handled more effectively \cite{baytas2017patient,che2018recurrent}. These methods established time intervals and missingness patterns as informative components of clinical sequence modeling.

More recent sequence models seek stronger temporal representations for irregular or longitudinal data. ODE-RNN and Latent ODE formulate hidden states as continuous-time processes for irregularly sampled observations \cite{rubanova2019latent}, while transformer-based models improve long-range dependency modeling through attention \cite{vaswani2017attention}. ContiFormer combines continuous-time dynamics with attention for irregular sequence modeling \cite{chen2023contiformer}, and longitudinal medical imaging models such as LOMIA-T adapt temporal augmentation, cross-attention, and longitudinal fusion ideas to treatment-response prediction \cite{sun2024lomia}. In parallel, diffusion-based medical prediction models have been explored for conditional classification, denoising-based diagnosis, sequence-aware generation, multimodal prediction, and treatment-aware longitudinal synthesis \cite{han2022card,yang2023diffmic,yoon2023sadm,wu2024mmfusion,liu2025treatment}. Nevertheless, these methods mainly address temporal aggregation, missingness-aware encoding, attention-based dependence modeling, or future-state generation, without explicitly correcting abrupt trajectory shifts during continuous ophthalmic follow-up. This limits their ability to track treatment- or recurrence-related changes over long horizons and weakens the reliability of endpoint prognosis prediction.

\subsection{Neural Differential Equations and Prototype Learning}
Neural differential equations provide a principled way to model latent evolution in continuous time. Neural ODEs parameterize the derivative of a hidden state with a neural vector field and use numerical integration to obtain the evolved state, establishing a general continuous-depth modeling paradigm \cite{chen2018neural}. This idea has been adopted in medical image analysis for deformation, registration, and multi-scale representation learning \cite{wu2022nodeo,xu2021multi}. NCDEs further extend continuous-time modeling by using an interpolated observation path as the control signal, allowing irregularly sampled measurements to drive the latent trajectory throughout time rather than only through an initial condition \cite{kidger2020neural}. Longitudinal medical imaging work such as LaTiM shows that continuous-time disease progression modeling can scale to large real-world ophthalmic cohorts \cite{zeghlache2024latim}.

Prototype learning offers a complementary representation principle by organizing latent spaces around representative class anchors. Prototypical networks introduced class prototypes as metric-learning centers for few-shot recognition \cite{snell2017prototypical}, and later time-series prototype methods used learned prototypes to support interpretable sequence classification \cite{gee2019explaining}. Recent medical time-series prototype learning further emphasizes human-machine collaboration and clinically meaningful prototype refinement, indicating that prototypes can serve not only as visualization tools but also as controllable semantic anchors \cite{xie2024prototype}. However, existing continuous-time models lack explicit class-aware geometric constraints on the evolving latent trajectory, whereas prototype-based methods rely on discrete or static representation anchoring and therefore lack the continuous-time control mechanism required to model irregular observation-driven follow-up dynamics.

\begin{figure*}[t]
    \centering
    \includegraphics[width=\textwidth]{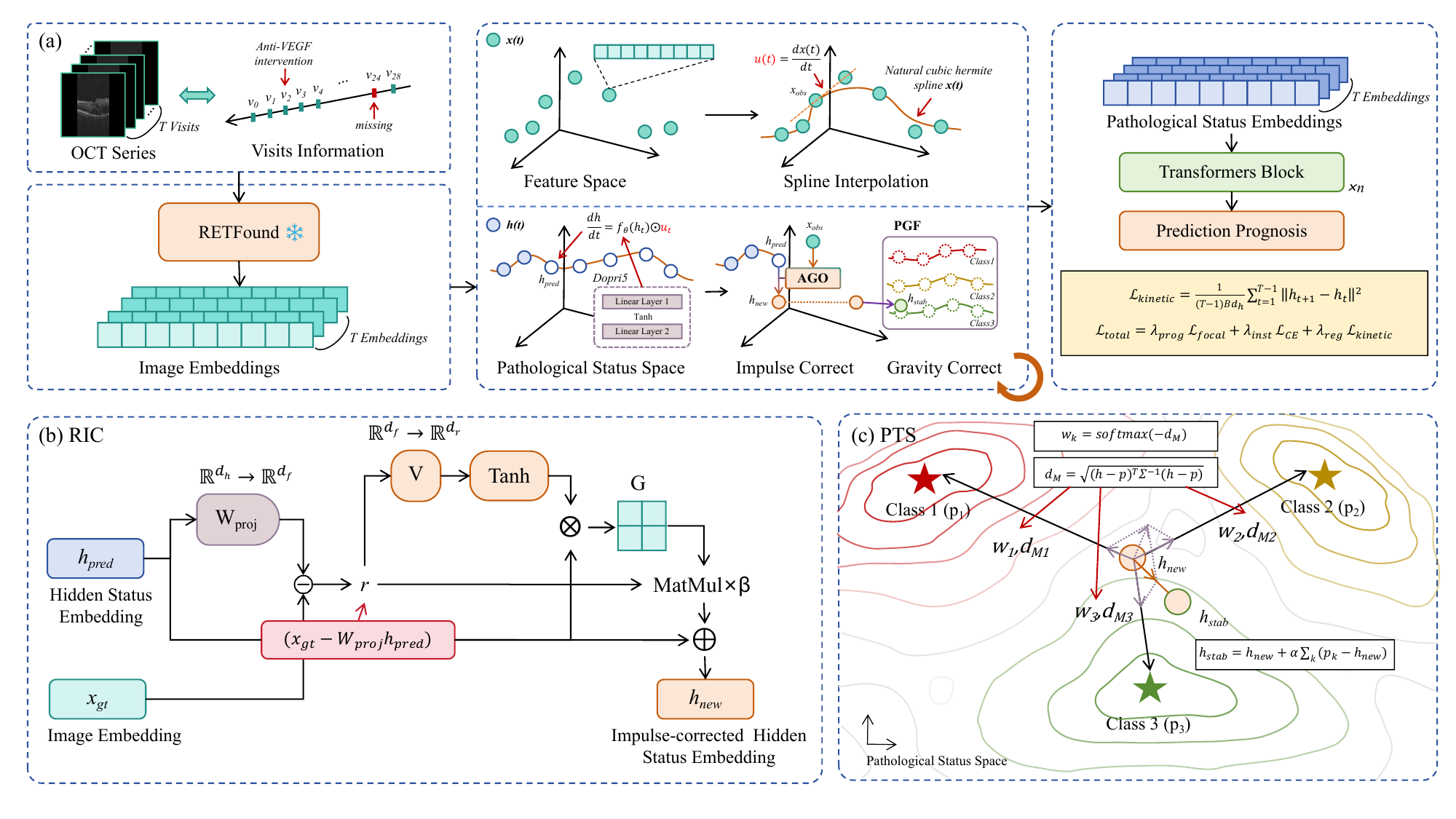}
    \caption{Overview of ImProNCDE. (a) Irregular OCT follow-up visits are encoded by RETFound and lifted into a continuous control path by natural cubic Hermite spline interpolation for NCDE-based latent disease progression modeling. (b) Residual Impulse Calibration (RIC) uses the residual mismatch between the predicted latent state and the newly observed imaging embedding to recalibrate abrupt pathological trajectory shifts at visit times. (c) Prototype-guided Trajectory Stabilizer (PTS) attracts corrected latent states toward learnable prognosis prototypes, reducing long-horizon drift and improving endpoint prognosis prediction.}
    \label{fig:flow}
\end{figure*}

\section{Methods}

\subsection{Overview}
Fig.~\ref{fig:flow} shows the overall ImProNCDE pipeline. We formulate longitudinal prognosis prediction as learning a patient-specific latent trajectory from irregular ophthalmic imaging visits, where the trajectory should represent both gradual disease evolution between visits and abrupt pathological shifts revealed during clinical follow-up. ImProNCDE follows this objective with three coupled components. First, a pretrained retinal encoder extracts visit-level imaging embeddings, and spline interpolation lifts these visit-level observations into a continuous control path. Second, an NCDE solver uses this control path to model latent disease evolution over unequal follow-up intervals. Third, RIC recalibrates the latent state when a newly observed imaging feature deviates from the continuous prediction, and PTS constrains the corrected trajectory toward learnable prognosis prototypes. The corrected latent sequence is finally aggregated by a transformer encoder to produce the endpoint prognosis.

\subsection{Spatial Semantic Embedding and Continuous Path Construction}
The first step transforms each ophthalmic imaging visit into a semantic representation that can serve as the input signal for continuous-time modeling. We use the retinal foundation encoder RETFound as the spatial backbone because it provides features that encode anatomical and pathological structures relevant to ophthalmic prognosis \cite{zhou2023foundation}. For each follow-up visit $t_i$, the input ophthalmic image $I_i$ is encoded into a semantic embedding $x_i$:
\begin{equation}
    x_i = \text{RETFound}(I_i) \in \mathbb{R}^{d_f}
\end{equation}
where $d_f = 1024$ is the feature dimension. The resulting sequence $\{(t_i,x_i)\}_{i=1}^{n}$ preserves the actual acquisition time of each visit rather than forcing all patients into a fixed temporal grid. Because ophthalmic measurements are sparse and irregularly sampled, these visit-level embeddings must be lifted to a continuous function before they can drive the differential equation solver. We apply natural cubic Hermite spline interpolation to construct a continuously differentiable trajectory $x(t)$ \cite{kidger2020neural}. The spline coefficients are determined by the observed feature values and estimated boundary derivatives within each interval $[t_i,t_{i+1}]$. The analytical derivative of the trajectory serves as the continuous control signal $u(t)$:
\begin{equation}
    u(t) = \frac{dx(t)}{dt} = b + 2c(t - t_i) + 3d(t - t_i)^2
\end{equation}
where $b, c, d$ are polynomial coefficients derived from the spline constraints. This construction converts irregular imaging visits into an observation-driven path, so the latent system can receive patient-specific temporal context continuously rather than only at the recorded visit times.

\subsection{Residual Impulse Calibration}
Standard NCDE dynamics evolve smoothly between observations, which is suitable for gradual disease progression but insufficient when clinical follow-up reveals a sudden pathological deviation. RIC recalibrates the latent trajectory at observed visits by converting the mismatch between prediction and observation into an impulse-like correction. At each visit time $t_{i+1}$, let $h_{pred}\in\mathbb{R}^{d_h}$ be the NCDE-predicted latent state and $x_{gt}\in\mathbb{R}^{d_f}$ be the newly observed imaging feature. RIC first computes the residual:
\begin{equation}
r = x_{gt} - W_{align}h_{pred},
\end{equation}
where $W_{align}\in\mathbb{R}^{d_f\times d_h}$ maps the predicted latent state to the imaging feature space, and $r\in\mathbb{R}^{d_f}$. A large residual indicates that the continuous trajectory has missed a follow-up trajectory deviation, such as an abrupt anatomical change or recurrent disease activity. The residual is then projected into a compact residual code:
\begin{equation}
e = Vr,
\end{equation}
where $V\in\mathbb{R}^{d_p\times d_f}$ and $e\in\mathbb{R}^{d_p}$, with $d_p=64$ in our implementation. To model the interaction between the current latent state and the residual evidence, RIC computes an outer-product interaction:
\begin{equation}
B = h_{pred}\otimes e,\quad B\in\mathbb{R}^{d_h\times d_p},
\end{equation}
where $\otimes$ denotes the vector outer product. This interaction is mapped by a learnable gain function $\psi(\cdot)$ into a residual-to-state gain matrix:
\begin{equation}
A = \psi(B),\quad A\in\mathbb{R}^{d_h\times d_f}.
\end{equation}
The impulse correction is then obtained by a matrix-vector product with the original residual:
\begin{equation}
\Delta h = Ar,\quad \Delta h\in\mathbb{R}^{d_h},
\end{equation}
and the hidden state is recalibrated as:
\begin{equation}
h_{new}=h_{pred}+\beta\Delta h,
\end{equation}
where $\beta$ is a learnable scaling factor initialized to a small value for training stability. The parameters $W_{align}$, $V$, and $\psi(\cdot)$ are shared across all visits and patients. This design preserves the NCDE trajectory as the default smooth evolution, while allowing the model to apply a visit-level correction when the observed imaging feature provides evidence that the predicted latent state has drifted from the actual disease status.

\subsection{Prototype-guided Trajectory Stabilizer}
RIC corrects visit-level deviations, while long-horizon prognosis still requires the latent trajectory to remain organized around outcome-relevant regions of the latent space. PTS addresses this issue by imposing a class-aware geometric constraint through learnable prototype vectors ${p_k}*{k=1}^K$, where each prototype represents an outcome-related latent anchor. After RIC correction, the proximity between the corrected state $h*{new}$ and each prototype is measured by a diagonal Mahalanobis distance \cite{mclachlan1999mahalanobis}:
\begin{equation}
d_M(h_{new}, p_k) = \sqrt{(h_{new} - p_k)^T \Sigma^{-1} (h_{new} - p_k)}
\end{equation}
where $\Sigma=\mathrm{diag}(\sigma_1^2,\ldots,\sigma_{d_h}^2)$ is a learnable diagonal covariance matrix shared across prototypes and visits. This diagonal metric does not model cross-dimensional correlations; instead, it learns dimension-wise scaling of the latent space and is more flexible than an isotropic Euclidean metric. The distance values are converted into normalized attraction weights with a softmax:
\begin{equation}
w_k = \frac{\exp(-d_M(h_{new}, p_k))}{\sum_{j=1}^K \exp(-d_M(h_{new}, p_j))}
\end{equation}
where $w_k$ represents the affinity between the corrected state and the $k$-th prototype. A prototype-guided force then pulls the latent state toward the weighted prototype barycenter:
\begin{equation}
h_{stabilized} = h_{new} + \alpha \sum_{k=1}^K w_k (p_k - h_{new})
\end{equation}
where $\alpha$ modulates the attraction strength. This stabilization regularizes each corrected visit state with prognosis-related anchors, reducing long-horizon latent drift and improving endpoint separability.

\subsection{Temporal Aggregation and Optimization Objective}
After NCDE evolution, RIC recalibration, and PTS stabilization, ImProNCDE obtains a corrected latent sequence $H=[h_1,h_2,\dots,h_n]$ aligned with the observed visits. We feed this sequence into a transformer encoder \cite{vaswani2017attention} with positional embeddings to aggregate cross-visit dependencies. The classification token is then passed through a linear prediction head to estimate the sequence-level endpoint prognosis.

The model is optimized with endpoint prognosis supervision. Let $y$ denote the sequence-level prognosis label and $\hat{y}$ denote the predicted class probability. The prognosis loss is defined as:
\begin{equation}
L_{prog}=-\sum_{c=1}^{C}\mathbb{I}(y=c)\log \hat{y}_c .
\end{equation}
This objective supervises only the final prognosis prediction of the whole longitudinal sequence and does not impose endpoint labels on intermediate visit states.

\subsection{Ethics Statement}
The private longitudinal ophthalmic imaging data used in this study were collected from two clinical centers of Shanghai General Hospital, namely Shanghai General Hospital Hongkou Branch and Shanghai General Hospital Songjiang Branch. This study was reviewed and approved by the Institutional Review Board of Shanghai General Hospital under the project “Prediction of Anti-VEGF Treatment Response and Follow-up Strategy for Diabetic Macular Edema Based on Multimodal Imaging” (IRB Approval No. YLK-2026-109; Protocol No. 20260126094702388). All procedures involving human participants were conducted in accordance with institutional ethical requirements and the Declaration of Helsinki. The requirement for signed written informed consent was waived by the Institutional Review Board of Shanghai General Hospital, and all data were de-identified before analysis.

\section{Experiments}

\subsection{Datasets}
We evaluate ImProNCDE on two public longitudinal ophthalmic imaging datasets, OCT4DME and GRAPE, and one in-house longitudinal ophthalmic imaging cohort, denoted as OURS. These datasets cover heterogeneous follow-up settings, image sources, and prognosis endpoints. All public datasets were converted into a unified patient-eye sequence format, where each sequence contains ordered visits, visit-level image paths, available clinical variables, and a sequence-level prognosis label. Unless otherwise specified, public-dataset experiments use sequence-level splits with cached visit features and a larger test protocol to stress generalization under class imbalance.

OCT4DME is a public diabetic macular edema dataset for treatment-response prediction from the APTOS 2021 Big Data Competition \cite{zhang2026predicting}. The dataset contains longitudinal ophthalmic images and treatment-response annotations for anti-VEGF therapy. We use the labeled training entries to construct patient-eye samples under a short-term longitudinal treatment-response setting, where pre-treatment and post-treatment follow-up information is paired according to the image filename codes. The prediction target is a binary endpoint indicating whether continued anti-VEGF injection is required after treatment. Image features are extracted using the same feature-caching protocol as the other datasets, and the cohort is evaluated under the same protocol.

GRAPE is a public longitudinal glaucoma dataset designed for visual-field progression analysis and glaucoma management \cite{huang2023grape}. It provides longitudinal follow-up records with ophthalmic images, visual-field measurements, structural retinal measurements, and clinical information. In our evaluation, we use the available ROI ophthalmic images and construct each patient-eye sample by grouping images named by subject, laterality, and visit index. Because GRAPE does not provide an explicit treatment timeline, all visit-level intervention indicators are set to zero, making it a natural-progression cohort rather than an intervention-driven cohort. The prediction target is binary visual-field progression status.

OURS is an in-house multi-center longitudinal ophthalmic imaging cohort collected from DR patients undergoing anti-VEGF therapy at the Shanghai General Hospital Hongkou Branch and Songjiang Branch. It contains 465 subjects with serial ophthalmic images acquired at irregular visit time points from week -1 (screening) to week 52 after treatment, with therapy interventions performed at week 4 or week 6. Each sequence is assigned a clinician-annotated three-class endpoint prognosis label: cured, not cured, or relapsed. We split the dataset into 372 training subjects and 93 test subjects using stratified sampling to preserve the class distribution across cohorts.

\subsection{Experiment Setup}
We design the experiments to evaluate both comparative effectiveness and component-level contribution. For quantitative comparison, we compare ImProNCDE with representative baselines from three groups. The first group includes traditional sequence models, RNN \cite{elman1990finding}, LSTM \cite{hochreiter1997long}, GRU \cite{cho2014learning}, and Transformer \cite{vaswani2017attention}, which aggregate visit-level features over time. The second group includes medical time-series backbones, Medformer \cite{wang2024medformer}, ViTST \cite{li2023time}, and RAINDROP \cite{zhang2021graph}, adapted to cached longitudinal imaging embeddings under the same feature-level protocol. The third group includes adapted diffusion-based prognosis backbones, CARD \cite{han2022card}, DiffMIC \cite{yang2023diffmic}, TrajDiff \cite{yoon2023sadm,liu2025treatment}, and MMFusion \cite{wu2024mmfusion}. These diffusion rows are implemented as feature-level prognosis baselines on cached longitudinal embeddings, rather than direct reproductions of the original image-generation pipelines. Public datasets are evaluated under the large-test cached-feature protocol, while the OURS cohort is evaluated under the in-house main protocol with auxiliary baseline rows for broader comparison.

For ablation analysis, we focus on the two proposed modules, RIC and PTS, using the in-house DR cohort (OURS). It contains longer follow-up horizons, more visit time points, and explicit anti-VEGF intervention records, making it more suitable for analyzing visit-level trajectory correction and long-horizon stabilization. The RIC ablations remove the full impulse correction, the adaptive impulse gate, or the residual alignment map to examine whether visit-level deviation correction requires each subcomponent. The PTS ablations remove prototype stabilization, replace learnable prototypes, or remove the Mahalanobis distance to evaluate whether class-aware latent anchors and feature-correlation-aware geometry contribute to long-horizon trajectory separation. All ablation variants are evaluated with macro precision, recall, and F1.

\subsection{Evaluation Metrics}
We use accuracy, precision, recall, F1 score, and the area under the receiver operating characteristic curve (AUC) to evaluate prognosis prediction. For multi-class evaluation, class-wise metrics are computed in a one-versus-rest manner and macro-averaged across classes. Per-class F1 scores are additionally reported in the prototype-number analysis to examine whether prototype design affects minority-class recognition.

\subsection{Implementation Details}
ImProNCDE is implemented in PyTorch with hidden-state dimension $d_h=128$. The neural vector field $f_{\theta}$ is a two-layer MLP with tanh activations. RIC uses a 64-dimensional residual encoder and an initial impulse scale $\beta=0.1$. PTS maintains $K=9$ learnable prototypes with a diagonal Mahalanobis covariance matrix and attraction strength $\alpha=0.1$, following the prototype-number analysis in Fig.~\ref{fig:prototype_ablation}. The prediction head is a two-layer transformer encoder. We train all models with Adam \cite{kingma2014adam} and cosine annealing for 100 epochs. All experiments are conducted on a single NVIDIA RTX A6000 GPU with 48GB memory size.

\begin{table*}[!t]
  \centering
  \small
  \setlength{\tabcolsep}{2.4pt}
  \renewcommand{\arraystretch}{1.03}
  \caption{Quantitative comparison on public and in-house longitudinal ophthalmic prognosis datasets. Metrics are reported as percentages, and bold values indicate the best result within each dataset.}
  \label{tab:main_comparison}
  \begin{tabularx}{\textwidth}{
      @{}p{1.6cm}
      p{4.1cm}
      *{5}{>{\centering\arraybackslash}X}
      @{}
  }
    \hline\noalign{\vskip 1pt}\hline
    \textbf{Dataset} & \textbf{Models} & \textbf{Acc.} & \textbf{Prec.} & \textbf{Recall} & \textbf{F1} & \textbf{AUC} \\
    \midrule
    \multirow{12}{*}{OCT4DME}
      & RNN & 71.56 & 71.30 & 73.30 & 70.85 & 82.38 \\
      & LSTM & 71.10 & 73.18 & 75.01 & 70.88 & 82.88 \\
      & GRU & 71.56 & 73.10 & 75.07 & 71.27 & 83.00 \\
      & Transformer & 71.56 & 77.70 & 78.01 & 71.56 & 82.98 \\
    \cmidrule(l){2-7}
      & Medformer & 72.48 & 75.60 & 77.25 & 72.36 & 82.78 \\
      & ViTST & 75.23 & 77.57 & 79.67 & 75.06 & 81.13 \\
      & RAINDROP & 64.68 & 32.34 & 50.00 & 39.28 & 69.01 \\
    \cmidrule(l){2-7}
      & CARD & 69.27 & 76.74 & 76.24 & 69.26 & 83.10 \\
      & DiffMIC & 69.72 & 76.92 & 76.60 & 69.72 & 83.38 \\
      & TrajDiff & 71.10 & 72.47 & 74.42 & 70.78 & 82.73 \\
      & MMFusion & 72.48 & 72.22 & 74.30 & 71.79 & 79.72 \\
    \cmidrule(l){2-7}
      & \textbf{ImProNCDE} & \textbf{77.98} & \textbf{79.58} & \textbf{82.09} & \textbf{77.76} & \textbf{84.92} \\
    \midrule
    \multirow{12}{*}{GRAPE}
      & RNN & 74.62 & 61.11 & 57.06 & 58.21 & 60.74 \\
      & LSTM & 71.79 & 49.80 & 49.71 & 49.23 & 61.76 \\
      & GRU & 76.92 & 52.27 & 52.65 & 52.37 & 64.12 \\
      & Transformer & 64.10 & 51.92 & 53.82 & 49.44 & 58.24 \\
    \cmidrule(l){2-7}
      & Medformer & 64.10 & 55.83 & 62.35 & 52.93 & 65.29 \\
      & ViTST & 69.23 & 57.69 & 65.29 & 56.67 & 59.41 \\
      & RAINDROP & 79.49 & 54.12 & 54.12 & 54.12 & 55.88 \\
    \cmidrule(l){2-7}
      & CARD & 74.36 & 56.11 & 59.71 & 56.47 & 59.41 \\
      & DiffMIC & 79.49 & 59.60 & 62.65 & 60.61 & 61.18 \\
      & TrajDiff & 76.92 & 57.66 & 61.18 & 58.46 & 64.12 \\
      & MMFusion & 74.36 & 56.11 & 59.71 & 56.47 & 63.53 \\
    \cmidrule(l){2-7}
      & \textbf{ImProNCDE} & \textbf{89.74} & \textbf{79.17} & \textbf{68.53} & \textbf{72.14} & \textbf{71.18} \\
    \midrule
    \multirow{12}{*}{\textbf{OURS}}
      & RNN & 65.78 & 65.23 & 64.52 & 63.74 & 83.49 \\
      & LSTM & 71.66 & 68.99 & 69.23 & 68.24 & 85.75 \\
      & GRU & 71.12 & 69.14 & 69.43 & 69.17 & 85.86 \\
      & Transformer & 70.05 & 71.30 & 69.85 & 69.75 & 85.35 \\
    \cmidrule(l){2-7}
      & Medformer & 66.31 & 63.87 & 64.51 & 63.79 & 82.35 \\
      & ViTST & 64.17 & 65.36 & 63.05 & 63.67 & 80.86 \\
      & RAINDROP & 61.50 & 59.17 & 59.72 & 59.37 & 69.21 \\
    \cmidrule(l){2-7}
      & CARD & 68.82 & 69.37 & 67.44 & 63.82 & 83.46 \\
      & DiffMIC & 73.12 & 72.41 & \textbf{72.22} & 71.22 & 87.65 \\
      & TrajDiff & 67.74 & 64.51 & 66.25 & 62.85 & 81.44 \\
      & MMFusion & 67.74 & 65.67 & 66.43 & 64.68 & 85.63 \\
    \cmidrule(l){2-7}
      & \textbf{ImProNCDE} & \textbf{73.30} & \textbf{73.60} & 72.00 & \textbf{71.50} & \textbf{89.56} \\
    \hline\noalign{\vskip 1pt}\hline
  \end{tabularx}
\end{table*}

\section{Results}

\subsection{Comparison With Existing Prognosis Models}
Table~\ref{tab:main_comparison} presents the quantitative comparison between ImProNCDE and existing prognosis models on three longitudinal ophthalmic datasets. Overall, ImProNCDE achieved the strongest performance across these datasets, ranking first in 12 of the 15 reported metric columns and improving the average score by 2.80 percentage points compared with the strongest competing result in each metric column. These three datasets represent different clinical follow-up scenarios. OURS is a multi-center treatment-intervention cohort with multiple follow-up visits, GRAPE is a multi-visit natural progression cohort without explicit medication intervention records, and OCT4DME is a short-term two-visit anti-VEGF treatment-response cohort. Therefore, the comparison evaluates ImProNCDE under multi-center interventional follow-up, natural disease progression, and short-term therapy-response settings.

On OURS, the adapted diffusion-based model DiffMIC was the strongest competing method, achieving 73.12\% accuracy, 72.41\% precision, 72.22\% recall, 71.22\% F1, and 87.65\% AUC. ImProNCDE achieved the best accuracy, precision, F1-score, and AUC, with 73.30\% accuracy, 73.60\% precision, 71.50\% F1, and 89.56\% AUC, while maintaining a comparable recall of 72.00\%. Compared with DiffMIC, ImProNCDE improved accuracy by 0.18 percentage points, precision by 1.19 percentage points, F1 by 0.28 percentage points, and AUC by 1.91 percentage points, with an average improvement of 0.67 percentage points across the five metrics. Among the traditional sequence models, Transformer obtained the highest F1-score of 69.75\%, and GRU obtained the highest AUC of 85.86\%. Among the medical time-series models, Medformer performed best, with 63.79\% F1 and 82.35\% AUC.

On GRAPE, medical time-series and diffusion-based models were the most competitive for recall-oriented prediction. ViTST achieved the highest recall and DiffMIC achieved the highest F1-score, with 65.29\% recall and 60.61\% F1. ImProNCDE achieved the highest accuracy, precision, recall, F1-score, and AUC, with 89.74\% accuracy, 79.17\% precision, 68.53\% recall, 72.14\% F1 and 71.18\% AUC. Compared with the strongest competing values in each metric column, ImProNCDE improved accuracy by 10.25 percentage points, precision by 18.06 percentage points, recall by 3.24 percentage points, F1 by 11.53 percentage points, and AUC by 5.89 percentage points. The average difference across the five metrics was +9.79 percentage points. Among the medical time-series models, ViTST obtained the highest recall of 65.29\%, and Medformer obtained the highest AUC of 65.29\%. Among the diffusion-based models, DiffMIC achieved the highest F1-score of 60.61\%.

\begin{table*}[!t]
  \centering
  \small
  \setlength{\tabcolsep}{3.4pt}
  \renewcommand{\arraystretch}{1.12}
  \caption{Ablation studies of RIC and PTS on the in-house DR cohort (OURS). Metrics are reported as macro percentages.}
  \label{tab:ablation}
  \begin{tabularx}{\textwidth}{
    @{}
    >{\raggedright\arraybackslash}p{0.18\textwidth}
    *{3}{>{\centering\arraybackslash}X}
    @{\hspace{8pt}}
    >{\raggedright\arraybackslash}p{0.20\textwidth}
    *{3}{>{\centering\arraybackslash}X}
    @{}
  }
    \hline\noalign{\vskip 1pt}\hline
    \multicolumn{4}{c}{\textbf{RIC Ablation}} &
    \multicolumn{4}{c}{\textbf{PTS Ablation}} \\
    \cmidrule(r){1-4}\cmidrule(l){5-8}
    \textbf{Models} & \textbf{Prec.} & \textbf{Recall} & \textbf{F1}
    & \textbf{Models} & \textbf{Prec.} & \textbf{Recall} & \textbf{F1} \\
    \midrule
    ImProNCDE
    & \textbf{75.12} & 72.00 & \textbf{71.50}
    & ImProNCDE
    & \textbf{75.12} & \textbf{72.00} & \textbf{71.50} \\

    \hspace*{0.8em}w/o Impulse
    & 36.95 & 36.83 & 35.89
    & \hspace*{0.8em}w/o Prototype
    & 39.62 & 54.76 & 44.61 \\

    \hspace*{0.8em}w/o Adaptive Gate
    & 73.60 & \textbf{73.49} & 71.24
    & \hspace*{0.8em}w/o Learnable Proto
    & 58.39 & 49.84 & 49.74 \\

    \hspace*{0.8em}w/o Residual Align.
    & 39.77 & 40.44 & 38.94
    & \hspace*{0.8em}w/o Mahalanobis
    & 60.34 & 60.05 & 59.42 \\
    \hline\noalign{\vskip 1pt}\hline
\end{tabularx}
\end{table*}

On OCT4DME, the medical time-series backbone ViTST was the strongest competing model for accuracy and F1-score, while DiffMIC achieved the highest competing AUC. ImProNCDE achieved the highest performance across all five metrics, with 77.98\% accuracy, 79.58\% precision, 82.09\% recall, 77.76\% F1, and 84.92\% AUC. Compared with the strongest non-ImProNCDE result in each metric column, ImProNCDE improved accuracy by 2.75 percentage points, precision by 1.88 percentage points, recall by 2.42 percentage points, F1 by 2.70 percentage points, and AUC by 1.54 percentage points, corresponding to an average improvement of 2.26 percentage points. Among the traditional sequence models, Transformer obtained the highest F1-score of 71.56\%, and GRU obtained the highest AUC of 83.00\%. Among the diffusion-based models, MMFusion reached 71.79\% F1, and DiffMIC reached 83.38\% AUC.

\begin{figure}[!t]
  \centering
  \includegraphics[width=\columnwidth]{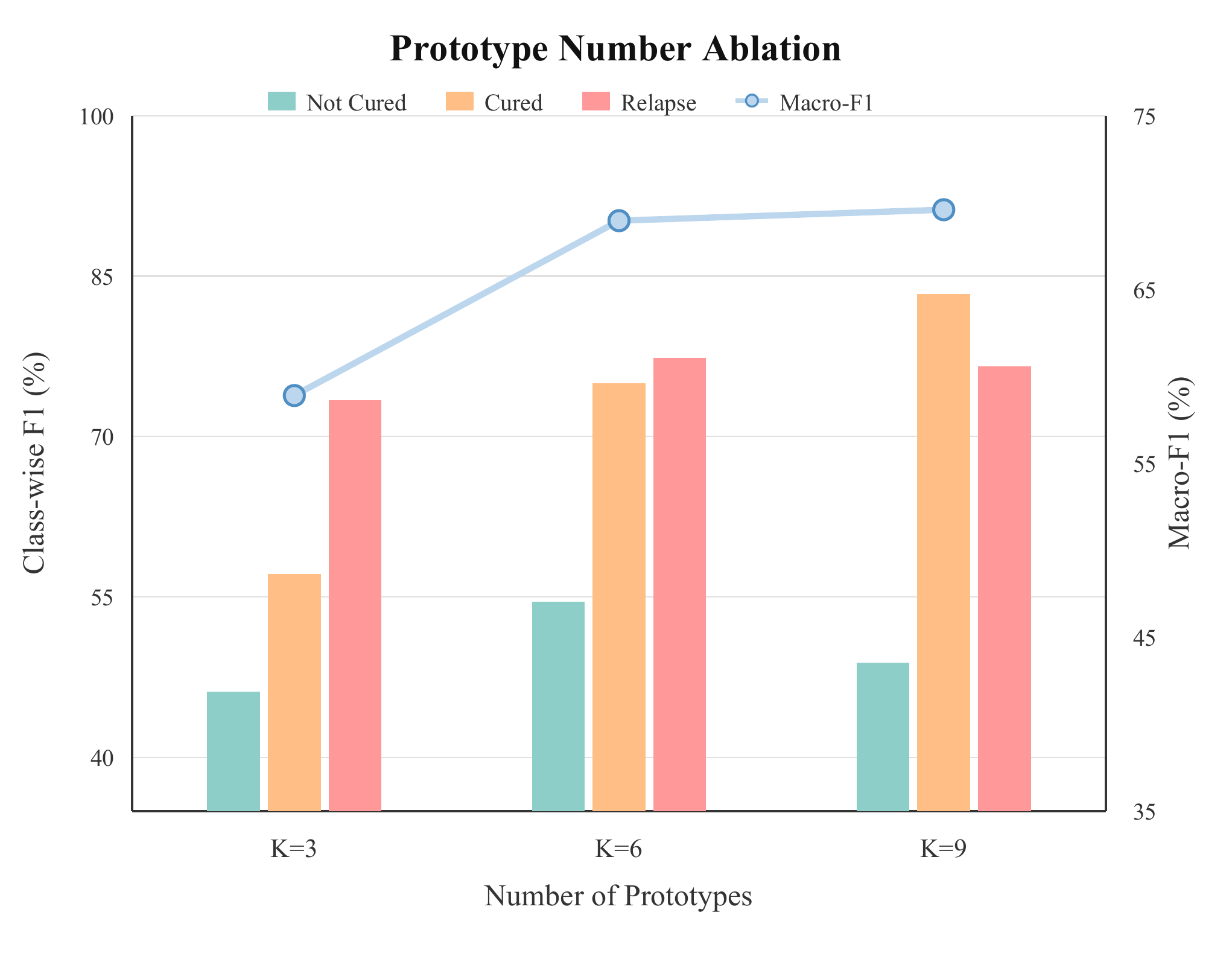}
  \caption{Prototype-number ablation on the in-house DR cohort (OURS). Bars report class-wise F1 scores, and the line reports macro-F1 under different numbers of learnable prototypes.}
  \label{fig:prototype_ablation}
\end{figure}

\begin{figure*}[!t]
  \centering
  \includegraphics[width=\textwidth]{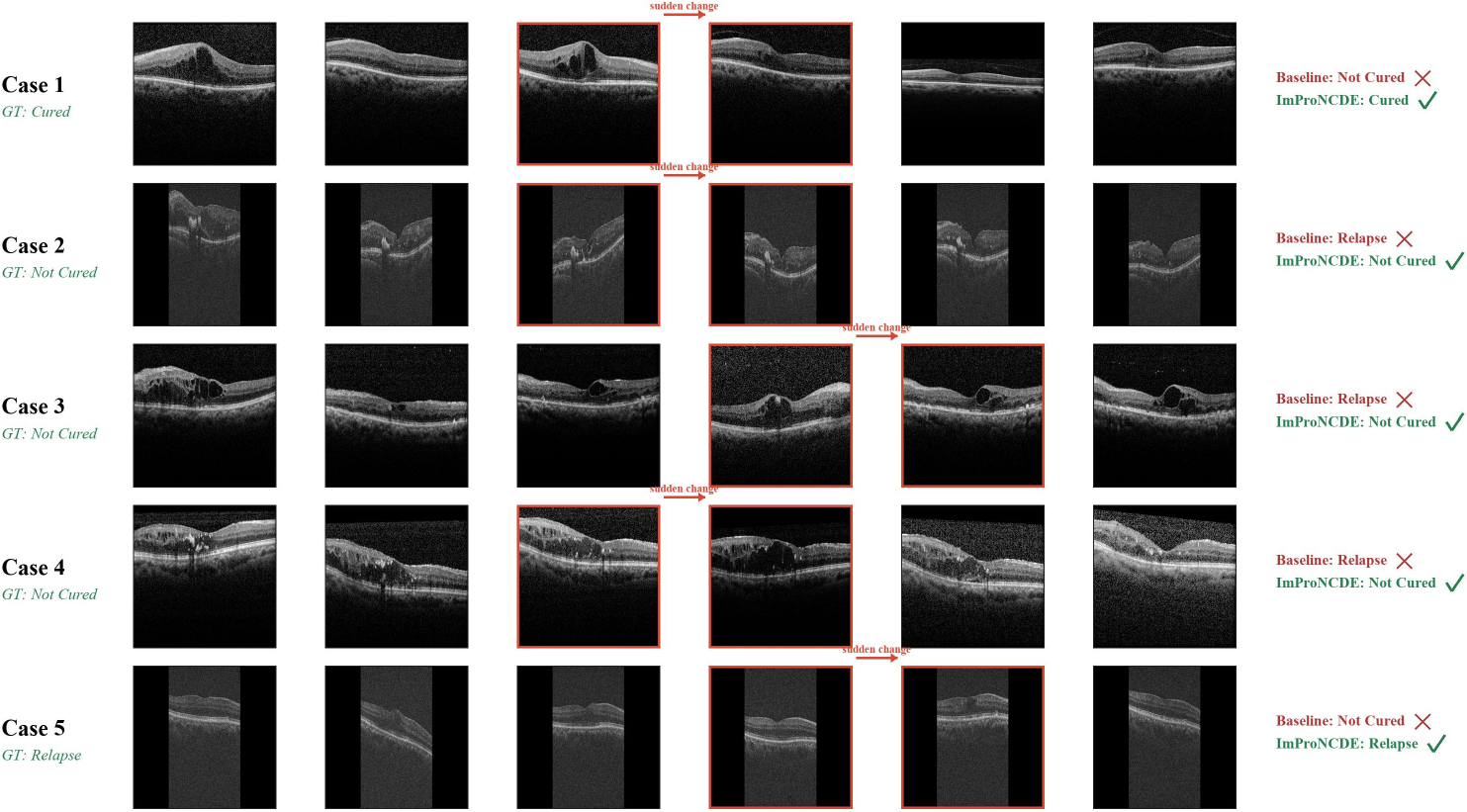}
  \caption{Case study on the in-house DR cohort (OURS). The montage shows representative longitudinal cases where abrupt pathological trajectory changes or long-term feature drift make the static first-visit baseline fail, whereas ImProNCDE predicts the correct prognosis by using longitudinal trajectory evidence.}
  \label{fig:case_study}
\end{figure*}

\begin{figure}[!t]
  \centering
  \includegraphics[width=0.66\columnwidth]{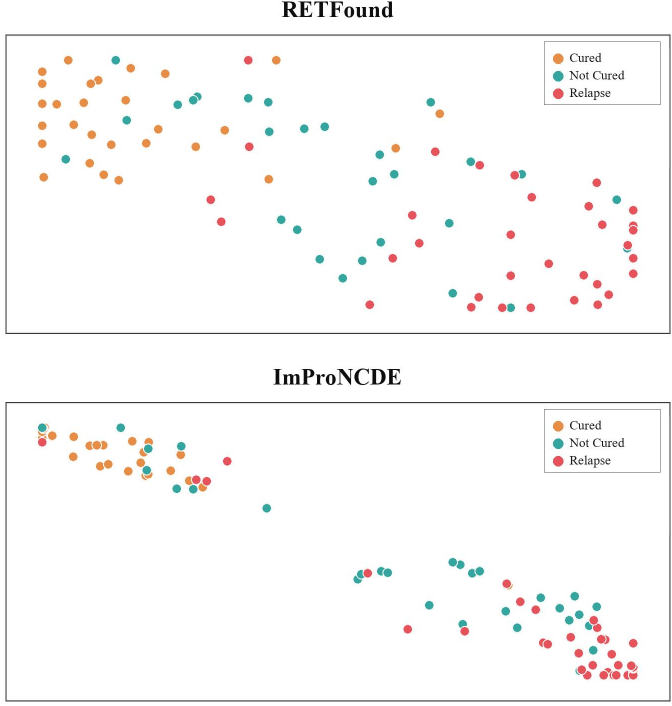}
  \caption{Prototype-guided representation clustering on the in-house DR test split. t-SNE visualization shows that ImProNCDE forms clearer prognosis-related clusters than RETFound-only features.}
  \label{fig:prototype_clustering}
\end{figure}

\subsection{Ablation Study}
Table~\ref{tab:ablation} presents the ablation results of RIC and PTS on the in-house DR cohort. The full ImProNCDE model achieved 75.12\% precision, 72.00\% recall, and 71.50\% F1. Removing the full impulse mechanism caused the largest performance decrease in the RIC group, reducing F1 from 71.50\% to 35.89\%. Removing the residual alignment map also substantially reduced performance, with 39.77\% precision, 40.44\% recall, and 38.94\% F1. In contrast, removing the adaptive gate produced a smaller change, with 73.60\% precision, 73.49\% recall, and 71.24\% F1. These results show that the residual-based impulse correction and residual alignment are the major contributors within RIC.

The PTS ablation showed a similar performance pattern. Removing prototype stabilization reduced F1 to 44.61\%, and replacing the learnable prototypes reduced F1 to 49.74\%. Removing the Mahalanobis metric achieved the strongest result among the PTS ablation variants, with 60.34\% precision, 60.05\% recall, and 59.42\% F1, but it still remained below the full model by 12.08 percentage points in F1. Overall, the ablation results demonstrate that both visit-level residual correction and prototype-guided stabilization are necessary for maintaining robust prognosis prediction on the in-house cohort.

\subsection{Prototype Number Analysis}
Fig.~\ref{fig:prototype_ablation} presents the class-wise F1-scores and macro-F1 under different numbers of learnable prototypes. From the macro-F1 curve, increasing the prototype number from $K=3$ to $K=6$ led to a clear improvement, with macro-F1 increasing from 58.90\% to 68.96\%. When the prototype number further increased to $K=9$, the macro-F1 only increased slightly to 69.59\%, showing that the overall performance gain became much slower after $K=6$. A similar trend can be observed in the class-wise results. The cured class increased from 57.14\% at $K=3$ to 75.00\% at $K=6$, and the not-cured class increased from 46.15\% to 54.55\%. When $K$ increased from 6 to 9, the not-cured class changed modestly from 54.55\% to 48.89\%, while the cured class continued to increase to 83.33\%. The relapse class remained relatively stable after $K=6$, with 77.33\% at $K=6$ and 76.54\% at $K=9$. These results show that increasing the prototype number from 3 to 6 provides the main performance gain, while adding more prototypes mainly brings limited additional improvement.

\subsection{Longitudinal Case Study}
Fig.~\ref{fig:case_study} shows representative cases from the in-house DR cohort where ImProNCDE captured abrupt pathological or morphological changes during follow-up. These examples cover several types of visit-level changes that are difficult to describe from the first visit alone. In Case 1, ImProNCDE captured the recovery of macular thickness between the third and fourth follow-up visits, where the retinal morphology changed from edematous thickening to an almost normal macular appearance. In Case 2, the model responded to the morphological change caused by OCT image acquisition, where the retinal profile changed from an upper-right oblique pattern to a lower-left oblique pattern between two visits. In Case 3, ImProNCDE captured a spatial shift in the morphology of macular edema, indicating that the disease-related abnormality changed its location during follow-up. In Cases 4 and 5, the model captured the deterioration process of macular thickness, with a clear increase in retinal thickening over time. These cases show that ImProNCDE can perceive different forms of abrupt pathological-state changes, including edema resolution, acquisition-related morphology variation, spatial displacement of edema, and progressive macular thickening.

\subsection{Latent Representation Analysis}
Fig.~\ref{fig:prototype_clustering} compares pooled RETFound-only features with the ImProNCDE prognostic latent representations on the in-house test split. In the RETFound-only feature space, not-cured and relapse samples showed substantial overlap. In the ImProNCDE latent space, samples from the same prognosis category formed more compact clusters, and the three prognosis categories became more separated. The quantitative clustering metrics were consistent with the visualization. The silhouette score increased from 0.123 to 0.208, K-means ARI increased from 0.241 to 0.372, and K-means NMI increased from 0.287 to 0.369. These results indicate that ImProNCDE improves the organisation of the latent representation beyond the original image embeddings.

\section{Discussion}
\subsection{Comparison With Existing Prognosis Models}
Sequence models are competitive when the main information lies in the temporal order of observed visits, medical time-series models are effective when follow-up signals can be represented as regular temporal patterns, and diffusion-based models are strong when the task benefits from modeling treatment-related future states. ImProNCDE improves over these methods by combining continuous-time trajectory modeling, visit-level residual correction, and prototype-guided latent stabilization. This design allows the model to adapt to multi-center interventional follow-up, natural progression, and short-term therapy-response settings without assuming that all datasets share the same temporal structure.

On OURS, the strongest competing method was DiffMIC, an adapted diffusion-based prognosis model. This result is reasonable because OURS is a multi-center cohort with medication intervention and multiple follow-up visits, where the final prognosis is strongly affected by treatment response and subsequent pathological changes. Diffusion-based models can be competitive in this setting because they are designed to model future-state uncertainty and treatment-related progression patterns. However, feature-level diffusion baselines still infer prognosis through a learned future representation, and they do not explicitly correct the latent state when a new visit reveals a sudden change. ImProNCDE addresses this limitation by using NCDE dynamics to model the continuous component of disease evolution and RIC to recalibrate the trajectory at each observed visit. This explains why ImProNCDE further improved F1 and AUC over DiffMIC while maintaining comparable recall.

On GRAPE, diffusion-based models were the strongest competing group, with DiffMIC achieving the best overall baseline performance. GRAPE is a multi-visit natural progression cohort without explicit medication intervention records, and its temporal changes are more likely to reflect gradual disease evolution rather than treatment-induced abrupt responses. In this setting, diffusion-based models can perform well because they are able to model the distribution of disease trajectories and better fit continuous progression patterns under non-intervention conditions. Notably, ImProNCDE also maintained strong performance in this setting, indicating that its continuous-time modeling remains effective for non-intervention longitudinal data dominated by gradual disease progression. ImProNCDE achieved the best accuracy, precision, recall, F1-score, and AUC on GRAPE.

On OCT4DME, the strongest competing model was ViTST for accuracy and F1-score, while DiffMIC achieved the highest competing AUC. OCT4DME is a short-term two-visit anti-VEGF treatment-response cohort, so the prognostic signal is concentrated in the difference between the two visits. Time-series backbones can perform well in this setting because the temporal structure is simple and the model mainly needs to compare pre- and post-response imaging features. Diffusion-based models can also be competitive because treatment response can be framed as a short-term state transition. ImProNCDE achieved the best results across all five metrics because it treats the two-visit change as a controlled trajectory rather than a static feature difference. Even in this short-term setting, the residual correction mechanism helps the model identify whether the observed follow-up feature deviates from the predicted latent state, while PTS stabilizes the endpoint representation for binary response prediction.

\subsection{Ablation Study}
The ablation study demonstrates that RIC is the key component for handling abrupt visit-level deviations. Removing the full impulse mechanism caused the largest drop in F1-score, and removing the residual alignment map produced a similarly severe degradation. This suggests that the model depends on the residual between the predicted latent state and the newly observed imaging feature to detect whether the continuous trajectory has drifted from the actual disease status. The adaptive gate had a smaller effect, which indicates that the gate mainly regulates the strength of the residual correction rather than providing the correction signal itself.

PTS provides a complementary contribution by organising the corrected latent states around prognosis-related anchors. Without prototype stabilization, the endpoint representations are more likely to scatter after repeated NCDE integration and RIC updates. Replacing learnable prototypes or removing the Mahalanobis metric also reduced performance, showing that both adaptive prototype locations and dimension-wise latent scaling are useful for constructing a stable prognosis space. Unlike a simple classification head that only maps the final representation to a label, PTS acts directly on the latent trajectory and constrains the representation before temporal aggregation. This explains why the full model achieved the most balanced performance across precision, recall, and F1-score.

\subsection{Prototype Number and Representation Analysis}
The prototype-number experiment indicates that each prognosis class requires more than one latent anchor, but excessive prototype subdivision brings limited additional benefit. When $K=3$, the model effectively allocates one prototype to each of the three prognosis classes. This setting is too limited for the in-house DR cohort because patients with the same endpoint can still follow different longitudinal patterns, such as rapid recovery, delayed response, persistent activity, or relapse after temporary remission. Increasing $K$ from 3 to 6 provides approximately two prototypes for each class, which is sufficient to cover the main intra-class trajectory variations and leads to the major macro-F1 improvement. When $K$ further increases from 6 to 9, each class can be represented by more prototypes, but the performance gain becomes much smaller. This trend suggests that two prototypes per class already capture the dominant prognosis subpatterns, while additional prototypes mainly refine local partitions rather than changing the overall class structure.

The t-SNE analysis further supports this interpretation. RETFound features provide general ophthalmic image representations, but they are not explicitly organised by longitudinal prognosis. As a result, not-cured and relapse samples still overlap in the original feature space. After NCDE evolution, RIC correction, and PTS stabilization, the latent representations become more compact within each prognosis category and more separated across categories. The improvements in silhouette score, K-means ARI, and K-means NMI show that ImProNCDE changes the structure of the representation space rather than only improving the final classifier.

\subsection{Longitudinal Case Study}
The case study illustrates that ImProNCDE can capture different types of follow-up changes rather than relying only on the baseline image. These changes include pathological deterioration, pathological alleviation, spatial migration of lesions, and morphology variation caused by OCT acquisition angle. In Case 1, the model captured the alleviation of macular edema, where the macular thickness recovered from an edematous state to an almost normal appearance. In Cases 4 and 5, the model captured the opposite direction of change, where macular thickness clearly increased during follow-up. These examples show that ImProNCDE is sensitive to both improvement and deterioration of pathological states.

The case study also shows that not all abrupt changes are simple increases or decreases in lesion severity. In Case 3, macular edema shifted spatially, indicating that the disease-related abnormality changed its location over time. In Case 2, the major change came from OCT acquisition morphology, where the retinal profile changed from an upper-right oblique pattern to a lower-left oblique pattern. Such acquisition-related variation may disturb a model that simply accumulates visit features, because the appearance change does not necessarily correspond to true pathological progression. ImProNCDE addresses these situations by comparing the predicted latent state with the newly observed imaging feature at each visit. The resulting residual allows the model to recalibrate the latent trajectory when clinically meaningful changes occur and to reduce the influence of less reliable morphology variations through prototype-guided stabilization. Therefore, the qualitative results support that ImProNCDE can perceive and correct multiple forms of abrupt follow-up changes.

\subsection{Limitation and Future Work}
Several limitations remain in this study. First, the current evaluation is still limited to ophthalmology-related diseases. Although longitudinal ophthalmic imaging provides a representative setting for irregular follow-up and treatment response prediction, disease progression in other organs may involve different imaging modalities, anatomical structures, and temporal patterns. Therefore, the generalisability of ImProNCDE beyond ophthalmic prognosis prediction still requires further validation. Second, the current framework mainly focuses on pathological trajectory changes associated with pharmacological injection, especially anti-VEGF-related treatment response. Other clinical interventions, such as surgery, radiotherapy, device-assisted treatment, or lifestyle-based management, may induce different types of abrupt or delayed pathological changes, which are not fully covered by the current datasets. Third, ImProNCDE is currently implemented as a task-specific prognosis framework, and it has not yet been developed into a plug-and-play temporal enhancement module for broader pretrained medical imaging models.

Future work will focus on three directions. We will extend the evaluation to longitudinal datasets from more organ systems to examine whether the proposed trajectory correction mechanism generalises beyond ophthalmology. We will also include datasets with more diverse intervention types to evaluate whether residual impulse calibration can model broader intervention-related trajectory shifts. Finally, we aim to reformulate the impulse correction and prototype stabilization components as a lightweight plug-and-play module that requires only limited training and can be integrated with pretrained medical imaging models to improve their temporal modelling capability.

\section{Conclusion}
This study introduced ImProNCDE, an impulse-corrected neural controlled differential equation framework with prototype learning for longitudinal ophthalmic prognosis prediction. The proposed method models irregular follow-up trajectories in continuous time and uses Residual Impulse Calibration to correct latent states when newly observed images reveal abrupt pathological or morphological changes. It further incorporates a Prototype-guided Trajectory Stabilizer to organise corrected trajectories around prognosis-related latent anchors and improve endpoint separability. Experiments on public and in-house ophthalmic datasets showed that ImProNCDE consistently outperformed traditional sequence models, medical time-series backbones, and adapted diffusion-based prognosis models. The ablation study and qualitative analyses further validated the contributions of residual impulse correction and prototype-guided stabilization. These results demonstrate that ImProNCDE provides an effective framework for modelling sparse, irregular, and clinically dynamic ophthalmic follow-up sequences, while future work will extend it to larger external cohorts with richer treatment records.

\section*{References}

\bibliographystyle{IEEEtran}
\bibliography{references}

\end{document}